\documentclass[%
reprint,
 superscriptaddress,
 nofootinbib,
 amsmath,amssymb,
 aps,
 ]{revtex4-1}

\usepackage{graphicx}
\usepackage{hhline}
\usepackage{xcolor}
\usepackage{hyperref}
\usepackage{placeins}
\usepackage{adjustbox}

\graphicspath{ {./figures/} }
\DeclareUnicodeCharacter{2009}{\,}

\begin{document}

\bibliographystyle{plain}

\pagenumbering{arabic}

\title[]{Coalescing Compact Binary Parameter Estimation with Gravitational Waves in the Presence of non-Gaussian Transient Noise}
\author{Yannick Lecoeuche}
\email{ylecoeuc@student.ubc.ca}
\affiliation{
Department of Physics \& Astronomy, University of British Columbia, Vancouver, BC V6T 1Z1, Canada
}
\author{Jess McIver}
\affiliation{
Department of Physics \& Astronomy, University of British Columbia, Vancouver, BC V6T 1Z1, Canada
}
\author{Alan M. Knee}
\affiliation{
Department of Physics, University of Michigan, Ann Arbor, MI 48109, USA
}
\affiliation{
Department of Physics \& Astronomy, University of British Columbia, Vancouver, BC V6T 1Z1, Canada
}
\author{Rhiannon Udall}
\affiliation{
Department of Physics \& Astronomy, University of British Columbia, Vancouver, BC V6T 1Z1, Canada
}
\affiliation{
TAPIR, California Institute of Technology, Pasadena, CA 91125, USA
}
\affiliation{
LIGO Laboratory, California Institute of Technology, Pasadena, California 91125, USA
}
\author{Katie Rink}
\affiliation{
Department of Physics, University of Texas at Austin, Austin, TX 78712, USA
}
\affiliation{
Weinberg Institute for Theoretical Physics, University of Texas at Austin, Austin, TX 78712, USA
}
\affiliation{
Department of Physics \& Astronomy, University of British Columbia, Vancouver, BC V6T 1Z1, Canada
}
\author{Sophie Hourihane}
\affiliation{
TAPIR, California Institute of Technology, Pasadena, CA 91125, USA
}
\affiliation{
LIGO Laboratory, California Institute of Technology, Pasadena, California 91125, USA
}
\author{Simona J. Miller}
\affiliation{
TAPIR, California Institute of Technology, Pasadena, CA 91125, USA
}
\affiliation{
LIGO Laboratory, California Institute of Technology, Pasadena, California 91125, USA
}
\author{Katerina Chatziioannou}
\email{kchatziioannou@caltech.edu}
\affiliation{
TAPIR, California Institute of Technology, Pasadena, CA 91125, USA
}
\affiliation{
LIGO Laboratory, California Institute of Technology, Pasadena, California 91125, USA
}
\author{TJ Massinger}
\author{Derek Davis}
\affiliation{
LIGO Laboratory, California Institute of Technology, Pasadena, California 91125, USA
}
\affiliation{
Department of Physics, University of Rhode Island, Kingston, RI 02881, USA
}

\begin{abstract}

Data from gravitational-wave (GW) detectors often contains a high rate of non-Gaussian transient noise, known as glitches. The parameters estimated from GW signals coinciding with detector glitches are occasionally biased away from their true values. During the first part of the fourth LIGO-Virgo-KAGRA (LVK) observing run, 29\% of GW candidates had overlapping or nearby glitches in one or more detectors. In the latter part of the fourth observation run, sensitivity improvements have increased the rates of GW detection. Consequently, scenarios in which GW signals and detector glitches overlap in time are more likely. In this study, we quantify shifts in inferred posterior distributions for short-duration compact binary coalescence GW signals interacting with common LIGO glitches as a function of time between the signal merger time and the glitch. We find statistically significant biases in parameter estimation for mass, spin, and sky position for ``blip'', ``thunder'', and ``fast-scattering'' glitches. Using these results, we provide estimates of what parameters are most affected by overlapping noise sources, as well as what constitutes a ``safe" time separation between a gravitational wave signal and a glitch, without requiring glitch subtraction for unbiased source property estimation. We find that in a majority of cases, all parameters are susceptible to significant bias due to glitch interference. Additionally, we find that glitches that occur within the time prior of the GW signal cause more extreme biases than glitches outside of the time prior.

\end{abstract}

\maketitle

\section{Introduction}

Since 2015, Advanced LIGO (Laser Interferometer Gravitational-wave Observatory)~\cite{ligo}, Advanced Virgo~\cite{virgo}, and KAGRA have reported a total of 218 likely gravitational-wave (GW) candidates, with 128 of them occurring during the first part of the fourth observing run (O4)~\cite{gwtc4_catalog}. Analyses of these GW signals provide important information on a range of topics, including binary merger rates~\cite{gwtc4_catalog}, compact object populations~\cite{populations_gwtc4}, and tests of general relativity~\cite{gr_tests_gwtc4}. LIGO and Virgo's ability to detect signals and accurately estimate their source properties is limited by several factors, including data quality~\cite{detchar_o4} and waveform uncertainty~\cite{PhysRevD.63.082005, Moore_2021, Owen_2023}. GW detectors are often subject to non-Gaussian transient noise, known as glitches, that adversely affect data quality and can bias estimated GW parameters. These biases can then propagate into subsequent studies, influencing their results and leading to potentially false astrophysical conclusions. This study focuses on the impact of glitches on parameter estimation of detected compact binary coalescences (CBCs).

Glitches come from a wide variety of sources\footnote{There were 23 classes of recurring glitch identified in strain channel data during the third observing run~\cite{soni}.}, many of which do not have a well understood physical cause~\cite{Davis_2022, Acernese_2023, Soni_2021}. In the first part of the fourth observing run, the LIGO Hanford Observatory (LHO) and the LIGO Livingston Observatory (LLO) detectors experienced glitches with signal-to-noise ratio greater than 6.5 at a rate of 77.2 and 53.7 glitches per hour~\cite{Soni_2025}, respectively, and the average rate of confident signal detections was 0.36 per day~\cite{gwtc4_catalog}. As detector upgrades progress, improvements in both sensitivity and duty cycle are expected to increase the rate of detected GW signals. However, lowering the noise floor may expose glitch sources that previously had gone undetected, and the introduction of new hardware to the detector can create new glitch sources or exacerbate the rate and severity of existing glitches. Due to these factors, the chance of GW signals coinciding in time with detector glitches is on average expected to increase over time, even as the total noise floor gets lower.

Our ability to extract information from GW detector data can be adversely affected by glitches coincident in time with GW signals. Powell~\cite{powell} injected binary black hole (BBH) signals over a set of glitches taken from the first observing run (O1), showing that chirp mass and luminosity distance posteriors are significantly biased for BBH injections directly overlapping and offset by 0.1 seconds from their blip glitch examples, particularly when the SNR of the glitch is larger than that of the GW signal. Another study by Macas et al (2022)~\cite{macas} investigated the effect the glitch types used in our study had on low-latency sky localization when coincident in time with four different CBC signals. Biases in low-latency localization can misdirect electromagnetic observatories away from the true sky location, causing them to lose valuable measurement time~\cite{pankow}. The study found that certain combinations of CBC signal, glitch type, and time offset between glitch peak and signal chirp could significantly offset sky localization for telescopes with field-of-view below 1 deg$^2$. Hourihane et al. (2025)~\cite{Hourihane_2025} found that glitches can significantly bias parameter estimation when energy from the glitch falls within the time prior of a CBC signal. Glitch characterization and mitigation are crucial to maximizing the science inferred from GW data.

Existing methods for removing glitches from strain data either rely on the fact that GW strain is coherent between multiple observatories while glitches are incoherent~\cite{bayeswave, Cornish_2021, Ghonge_2024, Chatziioannou_2021, Bondarescu_2023, Narola_2025, Macas_2024}, require auxiliary witnesses for noise subtraction~\cite{lin_subtraction}, or use other methods such as time-frequency area subtraction~\cite{pankow}. The most effective method uses a parameterized noise model that is capable of concurrently modeling glitch, GW signal, and background noise -- as employed in the code base \texttt{BayesWave}~\cite{bayeswave}. A drawback to this approach is that the \texttt{BayesWave} de-glitching process is computationally expensive and can take hours to days to complete, making it less useful for rapid interpretation of GW signal parameters. Additionally, the SNR of the post-subtraction residual glitch noise has a fundamental minimum limit, meaning it cannot perfectly remove biases from estimated parameters ~\cite{Udall_2025}. 
Recently, Udall et al. (2025)~\cite{Udall_2025} found that low-SNR noise such as that in post-subtraction residuals causes significant biases in GW parameter estimation, in particular for measured spin and precession. An improved understanding of when signal source properties are biased by coincident glitches would reduce the amount of de-glitching required, and could inform us on biases that might remain post-glitch subtraction~\cite{Hourihane_2025}.

Udall et al. (2025) found that joint inference of signal and glitch results in more accurate posterior distribution recovery than glitch subtraction followed by standard parameter estimation. Class-specific glitch models and joint inference methods have been the focus of several recent studies, allowing for improved inference for CBC signals with glitch overlap~\cite{Udall_2023, Ashton_2023, Malz_2025, Merritt_2021}. Our study utilizes the standard parameter estimation method \cite{bilby, bilby2}, which assumes only the presence of a CBC signal and Gaussian background noise.

In this study we select blip, thunder, and fast-scattering glitches, which are classified as being the most problematic\footnote{``Problematic'' in this case refers to the glitches having high rates of occurrence and a high degree of characteristic time-frequency overlap with CBC signals} to the accurate recovery of CBC sources from LIGO Livingston Observatory, which was the most sensitive detector during O3, and find examples of these glitches in strain channel data recorded during O3. We then inject a variety of simulated CBC signals (``injections'') into the strain data, incrementing in time to cover the duration of the glitch. We perform Bayesian parameter estimation for the set of injections using \texttt{Bilby}~\cite{bilby, bilby2}. Injecting into the most sensitive detector maximizes the impact of the glitch, giving a ``worst-case scenario'' for parameter estimation of our GW candidates when both LIGO detectors and the Virgo detector are online. We use a statistical metric, known as a ``cost function'', to quantify the difference in resulting posterior distributions from their injected values, relative to a set of baseline injections over glitch-free data.

This study builds off of previous work on biases in parameter estimation caused by glitch interference. The previously-mentioned study by Powell~\cite{powell} shares one glitch type with our study (blip glitches), and only includes data from the Hanford and Livingston observatories.\footnote{Advanced Virgo was not operational until the second observing run.} Our study extends the Powell study by using glitch data from a subsequent observing run (which has data from the Virgo detector), using a wider range of CBC signal masses, and examining a larger set of CBC parameters. Ghonge et al. (2024)~\cite{Ghonge_2024} investigated posterior biases for a simulated signal mimicking GW150914 injected over several glitch examples, and additionally investigated the biases after glitch-subtraction. Our study uses a wider variety of signal templates, two separate glitch types, and explores in more depth the possible interactions between signal and glitch. Another relevant study by Hourihane et. al~\cite{Hourihane_2025} measured posterior distribution change as a function of glitch proximity to a signal mimicking GW150914~\cite{gw150914}. Our study corroborates the results of Hourihane et. al with a larger variety of glitch types, a larger range of GW signals, and using real detector data.


The goal of this study is to investigate which GW signals are most susceptible to the most common O3 glitches across ten observable signal parameters, relating to the binary masses, their spins, and the sky position of the event. We also examine how close in time the signal chirp can be to
the beginning of each glitch before the signal parameters show significant biases. These susceptibility and ``safe" time separation measures provide guidelines as to which parameters of events with coincident glitches can be reliably trusted without subsequent glitch mitigation.

\section{Data}

The data used in this study were collected during the third observing run (O3), which can be accessed using the Gravitational Wave Open Science Center~\cite{Abbott_2023}. We use a three-detector network for our parameter estimation, consisting of LIGO Hanford, LIGO Livingston, and Virgo. Glitch examples are selected exclusively from LIGO Livingston data, due to it being the most sensitive detector during O3. Investigating glitches in the most-sensitive detector allows us to evaluate ``worst-case scenario'' posterior biases for the set of glitches presented in this section. The LIGO and Virgo observatories experience different glitch classes at varying levels of frequency and severity. Glitch types used in this study are representative of the types deemed most problematic at LLO during O3. The chosen glitch types and injected waveforms used in this study are also used in a study on low-latency sky localization for coincident GW signals and detector glitches by Macas et al. (2022)~\cite{macas}. Glitch examples are chosen such that LHO and Virgo strain data is glitch-free for the duration of the LLO glitch, as well as for a control period prior to the glitch.

We investigate biases in the following observable parameters, relating to the mass of the system, where subscripts $1$ and $2$ indicate the primary (more massive) and secondary (less massive) binary components.

\begin{itemize}
    \item $m_{1,2}$ - detector-frame mass of the binary component. Specified in $M_\odot$.
    \item $\mathcal{M}$ - detector-frame chirp mass of the system. Specified in $M_\odot$.
    \item $q$ - ratio of the secondary mass over the primary mass.
    \item $a_{1,2}$ - dimensionless spin of the binary component.
    \item $\chi_p$ - effective precessing spin value of the binary~\cite{Schmidt_2015}.
    \item $\chi_{\rm{eff}}$ - effective spin of the binary~\cite{Ajith_2011}. Mass-weighted projection of $a_1$ and $a_2$ onto the binary's orbital angular momentum.
    \item RA - right ascension of the system in the sky.
    \item DEC - declination of the system.
    \item $d_L$ - luminosity distance to the GW source. Specified in Mpc.
    \item $t_g$ - geocenter time of the event. Specified in seconds.
\end{itemize}

\subsection{Glitch examples}
\label{glitch examples}

\begin{figure}
    \centering
    \includegraphics[width=1.0\linewidth]{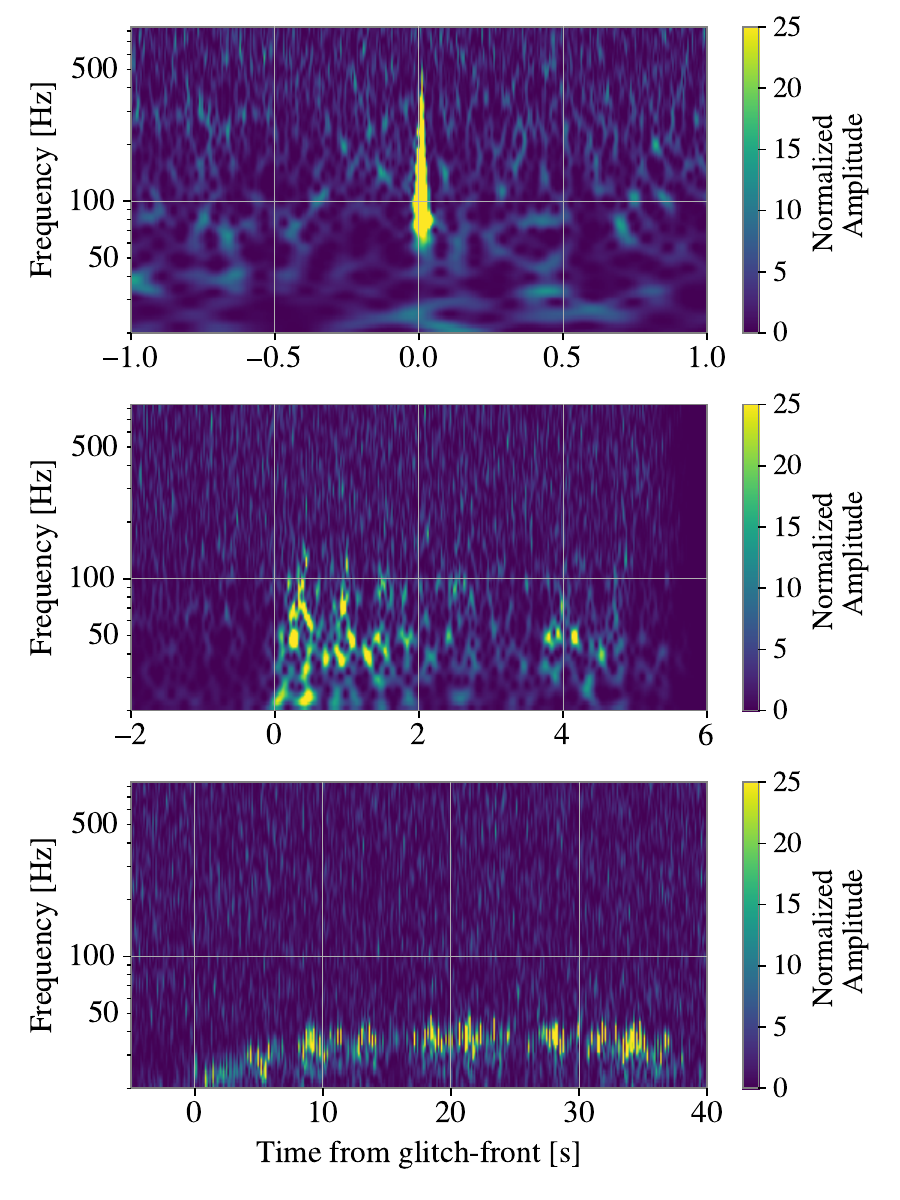}
    \caption{Glitch types selected for this study, chosen from LLO data as being the most problematic for recovery of CBC signals in the third observing run. Note the range in characteristic frequencies and timescales. Times on horizontal axis are given relative to the beginning of each glitch, as calculated in Section \ref{finding_glitchfronts}. \textit{Top}: Blip glitch, with sub-second characteristic timespan. \textit{Middle}: Thunder glitch, timespan on order of several seconds. \textit{Bottom}: Fast-scattering glitch, timespan on order of several minutes.}
    \label{glitch_types}
    \end{figure}

Three glitch types are chosen for this study: blip, thunder, and fast-scattering (see Figure \ref{glitch_types}). These glitch classes are chosen for being commonly seen in LLO strain channels during O3, and for introducing high SNR noise into the strain channels when they occur. 

\begin{itemize}
\item \textbf{Blip glitches} have sub-second timespans and have the widest frequency bandwidth of our three glitch classes. Only a small fraction of blip glitches have a known cause~\cite{miriam!}. Blip glitches can be mistaken for high mass binary black hole mergers by matched filtering searches, as a result of having similar time-frequency characteristics~\cite{davis}. 
\item \textbf{Thunder glitches} have timespans of several seconds, and generally cover frequencies under 200~Hz. Thunder glitches happen as a result of thunderstorms occurring near the Livingston detector acoustically coupling to the detector and causing light scattering noise~\cite{nutall}. 
\item \textbf{Fast-scattering glitches} are sub-second transients that have been found to occur in close succession for up to several minutes~\cite{Soni_2021}. They occur in the 20-60~Hz range, and are correlated with local ground motion~\cite{soni}. Fast scattering was the most frequent glitch type that occurred at Livingston during O3~\cite{gwtc3}, though between the third and fourth observing runs the frequency was reduced significantly by damping scattered light sources~\cite{Capote_2025}.
\end{itemize}

\begin{table}[]
\centering
\begin{tabular}{|l|l|l|l|}
\hline
                                                          \textbf{}  & \textbf{SNR}     \\
\hhline{|=|=|=|=|}
\multicolumn{1}{|l|}{Blip}            & 18.3, 12.2, 13.3   \\ \hline
\multicolumn{1}{|l|}{Thunder}             & 22.8, 13.3, 14.6      \\ \hline
\multicolumn{1}{|l|}{Fast-scattering}             & 18.6, 12.1, 12.0       \\ \hline
\end{tabular}
\caption{Peak SNRs of chosen glitch examples as estimated by spectrogram generation program Omicron. The maximum SNR among Q-transform tiles is reported.}
\label{glitch_snrs}
\end{table}

For each of these glitch types, three examples are selected from LLO strain data with the aid of a machine-learning glitch classifier called Gravity Spy~\cite{gspy, Zevin_2024, Wu_2025}. Chosen glitch examples are required to have glitch-free, Gaussian strain data for at least 64 seconds prior to the glitch in all three detectors. This ``clean" data prior to the glitch is used for a subset of our injections, from which we obtain a posterior distribution control group. We also use the time segment to measure our power spectral density (PSD) for waveform injections and parameter estimation. These examples are chosen to be representative of their glitch populations. The SNRs of our example events can be found in Table \ref{glitch_snrs}, as calculated by the time-frequency transform pipeline \texttt{Omicron}~\cite{Robinet_2020}. These peak SNRs are not calculated from models of the full glitch, but from single wavelet models fit to the highest-SNR tile in Q-transform space\footnote{The Q-transform space is logarithmically tiled in the time-frequency plane, with tiles having constant quality factor, or Q. Tile resolution in time-frequency space is a function of the quality-factor and the central frequency.}~\cite{Chatterji_2004}.

\subsection{Gravitational wave signals}
\label{gw_model}
    
     Three simulated GW signals are used for this study, based on the previously-detected binary black hole candidates: GW190521~\cite{gw190521}, GW150914~\cite{gw150914}, 
     and GW231123~\cite{gw231123}. The masses corresponding to these candidates span 29-137 $M_\odot$, allowing us to broadly inform on parameter estimation bias from coincident glitches and short-duration GW signals. The component source masses used for our simulated signals are the median values from the mass posterior distributions predicted by Bayesian inference (see Section \ref{PE}) for each of the CBC candidates listed below.
     
     \textbf{GW190521} is consistent with a BBH with component masses $85^{+21}_{-14}$~$M_\odot$ and $66^{+17}_{-18}$~$M_\odot$, forming a $\sim$140~$M_\odot$ black hole remnant. One or both of its mass components fall within the pair instability mass gap, and its remnant is the first directly detected intermediate mass black hole (IMBH).  Compared to lower-mass BBH mergers, GW190521 has a shorter duration of $\sim$0.1 seconds and spans a lower frequency range of 30-80~Hz. The signal is similar in time-frequency range to blip glitches and the sub-second transient components of fast-scattering glitches. In general the GW190521 mass parameters are well-constrained and the spin and sky parameters have relatively poor constraints.

     \textbf{GW150914} is consistent with a BBH with component masses $36^{+5}_{-4}$~$M_\odot$ and $29^{+4}_{-4}$~$M_\odot$, forming a $62^{+4}_{-4}$~$M_\odot$ black hole remnant. GW150914 was the first gravitational wave event detected by LIGO. The component masses fall within a known peak in the black hole primary mass populations, indicating that this is a relatively common type of CBC merger~\cite{populations}. It spans a frequency range of 35-250~Hz.
     
     \textbf{GW231123}, the most massive BBH observed to date, is consistent with component masses $137^{+22}_{-17}$~$M_\odot$ and $103^{+20}_{-52}$~$M_\odot$, forming a $225^{+26}_{-43}$~$M_\odot$ black hole remnant. The primary mass is estimated to have been within or above the pair-instability mass gap, while the secondary mass is estimated to be within or below the mass gap. This event suggests additional formation mechanisms beyond stellar collapse, such as hierarchical mergers~\cite{Kimball_2020}.

      The source mass parameters for the simulated signals in this study are chosen to model those of the original detections. 
      Our simulated CBC signals are modeled using waveform model \texttt{IMRPhenomXPHM}~\cite{xphm}, which is a model for quasi-circular precessing BBH mergers that includes subdominant harmonics. While the inferred posterior distributions reported in the GW231123 discovery paper were modeled with template \texttt{IMRPhenomXPHM}, the discovery papers for candidates GW190521 and GW150914 used alternate waveforms, the most similar to \texttt{IMRPhenomXPHM} being \texttt{IMRPhenomPv3HM}~\cite{Khan_2020} and \texttt{IMRPhenomPv2}~\cite{Khan_2019, Hannam_2014}, respectively. \texttt{IMRPhenomXPHM} has been shown to perform well relative to both models.
      
      We choose injected luminosity distance such that the recovered SNR remains above the minimum detection threshold for all injections (see Section \ref{signal_injection}). The right ascension and declination of each injection is chosen such that the Livingston observatory could detect them at maximum sensitivity at the time of the injection. Inclination is fixed at $\pi/4$, and the spins of each mass are uniformly set to zero for simplicity. Sampling rate, minimum frequency, and duration values are chosen to mirror those in the analysis of the CBC candidate we model.

  \begin{table}[]
    \begin{tabular}{|l|l|l|l|}
    \hline
        \textbf{} & \textbf{GW190521} & \textbf{GW150914} & \textbf{GW231123} \\
        \hhline{|=|=|=|=|}
       Blip & 10.2 - 12.5   & 10.9 - 13.4   & 10.9 - 11.5 \\ \hline
       Thunder & 13.0 - 14.9  & 12.8 - 15.1 & 13.5 - 15.0 \\ \hline
       Fast-scattering & 10.6 - 13.5  & 10.4 - 11.9 & 10.9 - 17.0  \\ \hline
    \end{tabular}
    \caption{List of ranges for injected CBC SNRs. Injection SNRs are chosen such that the lowest single template matched filter recovery SNR in the glitch-affected injections is close to but above the minimum detection threshold.}
\label{snrs}
\end{table}

\section{Methods}

 Methods in this study are separated into four subsections. In the first subsection we describe the process of injecting simulated signals around detector glitches. In the second subsection we explain how parameter estimation is performed on those signals using \texttt{Bilby}. In the third subsection we go over the statistical methods used to compare the posterior distributions for signals with coincident detector glitches to posteriors for a group of control injections into glitch-free data. Lastly we show our method for finding the beginning of each glitch example.

\subsection{Signal injection}
\label{signal_injection}

 For each of the nine glitch examples and three CBC models, we inject simulated signals into the data at intervals around the time of the glitch. We increment our injection points in time such that sixteen or more injections occur in the ``clean'' time prior to the glitch, to serve as a control group to understand the nominal variance of posteriors for each parameter of interest. We also use the injections prior to the glitch to evaluate the earliest point in time relative to the beginning of the glitch in which we see no significant bias in estimated posterior distributions. Injection merger times are chosen such that the majority of the energy of the signal passes through the entirety of the power in the glitch. Glitches within the same glitch category exhibit significant variations in morphology, timespan, and frequency, making it difficult to define injection time intervals that consistently sample the same interactions between signals and glitch examples~\cite{Udall_2025}.\footnote{A finer injection resolution would require significantly more computational power, where each Bilby inference uses approximately 142.5 CPU hours.} In each case a soft upper limit of 40 injections overlapping glitch data is used.

\begin{table}[]
    \begin{adjustbox}{width=\columnwidth,center}
    \begin{tabular}{|l|l|l|l|}
    \hline
        \textbf{Signal property} & \textbf{GW190521} & \textbf{GW150914} & \textbf{GW231123} \\
        \hhline{|=|=|=|=|}
       Minimum frequency  {[}Hz{]} & 11 & 20 & 10 \\ \hline
        Signal duration {[}s{]} & 8 & 8 & 8 \\ \hline
       Sampling rate {[}Hz{]} & 2048 & 2048 & 2048  \\ \hline
       Approximant: \texttt{IMRPhenom} & \small{\texttt{XPHM}} & \small{\texttt{XPHM}} & \small{\texttt{XPHM}} \\ \hline
    \end{tabular}
    \end{adjustbox}
    \caption{Relevant \texttt{Bilby} settings for CBC signals in this study. Minimum frequency, duration and sampling rate values for GW190521, GW190814, and GW231123 are taken from analysis of the original GW candidate. PSD windows span 64 seconds directly prior to the glitch.}
\label{priors}
\end{table}

 Each glitch and signal combination uses a single power spectral density measurement (PSD) for waveform injections and parameter estimation, measured from strain data taken directly prior to the glitch. PSDs are calculated using a median FFT-averaging on sixteen 8-second Tukey windows with 50\% overlap. As mentioned in Section \ref{gw_model}, the sky positions of our injected signals are optimally aligned with LLO at time of injection, resulting in the signal being detected with higher SNR in LLO strain data than the two other observatories. This creates a scenario in which our glitch example contaminates the strain data of a detector measuring the maximum possible fraction of the total signal in all three detectors (for a fixed network SNR). Parameter estimation biases from this will generally be larger than if the glitch occurs in either of the other detectors, or if the sky position of the event occurs in a less optimal location. This setup allows for a ``worst-case scenario'' evaluation of parameter estimation biases caused by overlapping glitches.

We subsequently utilize the matched-filter recovery search from \texttt{PyCBC} (see Appendix \hyperref[matchedfilter]{A}) to calculate the chi-square-weighted SNR of the injected template at each point of interaction~\cite{pycbc}. At different injection times, the excess noise due to the glitch will vary, changing the recovered SNR. If the excess noise for a particular injection causes the recovered network SNR to fall below the detection threshold of 8, the injection would not be detected and would therefore not be a representative example of glitch interference we might observe in the future. To understand maximum possible biases in parameter estimation due to glitches coincident with \textit{detectable} GW signals, the luminosity distance for each glitch type is chosen such that the lowest recovered chi-square-weighted network SNR across all injected waveforms/glitch examples is close to but never below the detection threshold. The range of CBC signal injections we perform is shown in Table \ref{snrs}. These SNRs range from $\sim$10-17, and are dependent on frequency and phase interactions between the signal and glitch. 

\subsection{Parameter estimation}
\label{PE}

 We perform parameter estimation on our post-injection strain data using Bayesian inference \cite{Thrane_2019, Veitch_2015} via the inference library \texttt{Bilby}~\cite{bilby}. Given strain data, $d$, and a waveform model, $h(\theta)$, where $\theta$ is the set of parameters describing the model, we can express the posterior probability distribution, $P(\theta|d)$ as:
\begin{equation}
    P(\theta|d) = \frac{L(d|\theta)\pi(\theta)}{Z(d)}
\end{equation}
where $L(d|\theta)$ is the likelihood function, $\pi(\theta)$ is the prior distribution, and $Z(d)$ is the evidence. We assume a stationary Gaussian noise likelihood, written in the frequency domain as:
\begin{equation}
    \ln L(d|\theta) = -\sum_k \frac{2|\tilde{d}_k-\tilde{h}_k(\theta)|^2}{S_{\rm n}(f_k)T}\,,
\end{equation}
where $\tilde{d}_k$ and $\tilde{h}_k$ are the Fourier transforms of the strain data and waveform model, respectively, $k$ is the frequency bin index, $S_{\rm n}(f_k)$ is the PSD of the detector noise, and $T$ is the duration of the analysis segment.
Lastly, the evidence serves to normalize the posterior, 
\begin{equation}
    Z(d) = \int\,L(d|\theta)\pi(\theta)\,{\rm d}(\theta)
\end{equation}
The inferred posterior distributions for the waveform parameters are the basis by which we evaluate the most probable parameter values. In later sections, we compare the estimated parameters for signals injected near detector glitches to estimated parameters for signals injected into Gaussian data. Chirp mass, mass ratio, geocenter time, and angular parameter priors are uniform, while luminosity distance prior is uniform in comoving volume. The component mass prior ranges act as constraints on the mass ratio and chirp mass priors. A subset of these prior ranges is listed in Table \ref{priors}, excluding geocenter time, which spans 0.1 seconds on either side of the injected chirp, and the angular parameters, which span the full parameter space.
The PSDs are set using the 64 seconds of Gaussian noise we selected for prior to each glitch. Sampling of the parameter space is performed with the \texttt{Dynesty} nested sampler~\cite{dynesty}\cite{bilby}\cite{bilby2}.

\begin{table}[]
\centering
\begin{tabular}{|l|l|l|l|}
\hline
                                                          \textbf{Parameters}  & \textbf{GW190521}     & \textbf{GW150914}    & \textbf{GW231123}     \\
\hhline{|=|=|=|=|}
\multicolumn{1}{|l|}{$m_{1}$ {[}$M_\odot${]}}             & 1-300          & 2-100       & 160-300    \\ \hline
\multicolumn{1}{|l|}{$m_{2}$ {[}$M_\odot${]}}             & 1-300          & 2-100       & 120-250      \\ \hline
\multicolumn{1}{|l|}{$q$}                                 & 0.02-1         & 0.02-1      & 0.02-1       \\ \hline
\multicolumn{1}{|l|}{$\mathcal{M}$ [$M_\odot$]}           & 1-150          & 3-120     & 130-230     \\ \hline
\multicolumn{1}{|l|}{$d_L$ [Mpc]}                         & 100-10,000     & 10-10,000    & 500-10,000      \\ \hline
\end{tabular}
\caption{Prior ranges used to estimate parameters for the three CBC signals used in the study. The geocenter time prior spans 0.1 seconds on either side of the chirp time. Prior ranges for angular parameters are isotropic on the unit sphere.}
\label{priors}
\end{table}

\begin{figure}
    \centering
    \includegraphics[width=1.0\linewidth]{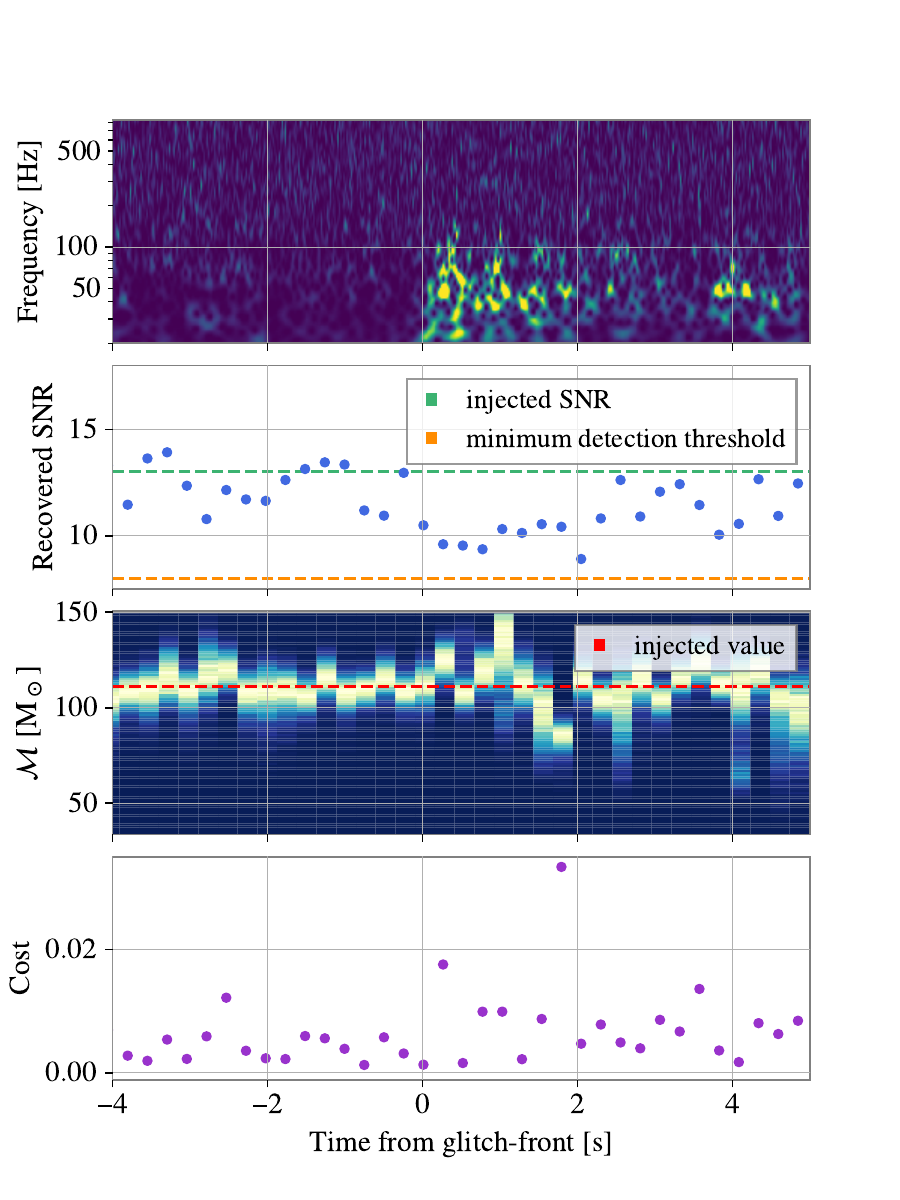}
    \caption{Ordered plots for the procedure of obtaining cost function values from GW signals injected around a glitch. Each vertically aligned set of data points in the bottom three plots corresponds to a single GW injection over the data, with the signal chirp occurring at that time. \textit{Top:} Omegascan of a blip glitch with no injections. \textit{Second from top:} Recovered SNR values using single template matched filtering (see Appendix \hyperref[matchedfilter]{A}). Green dotted line indicates the injected SNR of the signal, orange dotted line indicates the minimum detection threshold. \textit{Third from top:} 1D posteriors for the chirp mass ($\mathcal{M}$) for each injection, where lighter colors indicate higher probability density. Red dotted line indicates the injected value. \textit{Bottom:} Cost function output for each injection.}
    \label{multi-panel}
    \end{figure}

\subsection{Statistical Methods}
\label{stat_meth}

  In this section we set up a framework for comparing posterior distributions calculated from our injected signals. We measure the posterior distribution deviation relative to the true injection parameters via a cost function that relates the posterior median, standard deviation, and injected value:

\begin{equation}
    c = \frac{| \bar{x} - \mu |}{\sigma}
\end{equation}

\noindent where $\bar{x}$ is the median of the posterior distribution for a given parameter, $\mu$ is the injected value, and $\sigma$ is the standard deviation of the distribution. The function is most sensitive to posterior distributions with medians far from the injected value and low standard deviation, indicating high confidence in incorrect parameter space. The output of this function is normalized and equally weights posterior distributions above and below the injected value. If assuming the median of two distributions is identical, the ratio of their cost values gives the factor by which the standard deviation has changed. If assuming the standard deviations of two distributions are identical, the difference in cost function values gives the number of standard deviations separating the two medians.

Posterior distributions for CBC signals injected over ``clean'' backgrounds are expected to contain uncertainty due to fluctuations in the Gaussian noise background. To obtain a reference cost value for injections into glitch-free data, we average the cost function output for the first 16 time-separated injections within the 64 seconds of clean data prior to the glitch. The mean of the control cost function values is used as an indication of how well the parameters are estimated in the nominal case, and the standard deviation is used to determine the significance of cost function deviations that are potentially due to glitch interference.

While the cost function is a useful metric for determining changes in a distribution relative to a reference value, there is a cost function degeneracy between the median value and standard deviation. In order to better understand the type of bias being observed, we also compare the standard deviations of our control posterior distributions to subsequent injections outside of the control group.

Sky position posterior distributions are not described with the cost function, as they are often multimodal. Multimodal distributions can offset the median posterior values away from regions of high probability density and create standard deviation values that do not reflect the combined spread of individual peaks.
To calculate credible intervals for sky position we first created two-dimensional probability density functions (PDFs) over the right ascension and declination posteriors. PDFs are estimated using a kernel density estimator on the posterior samples. We then integrate over the PDF from regions of lowest density to highest density, until the bin containing the injected value is reached. The reported credible interval is the total integrated probability density for which the true value has more support, ranging from 0 to 1.

 \begin{figure}[!ht]
    \centering
    \includegraphics[width=0.85\linewidth]{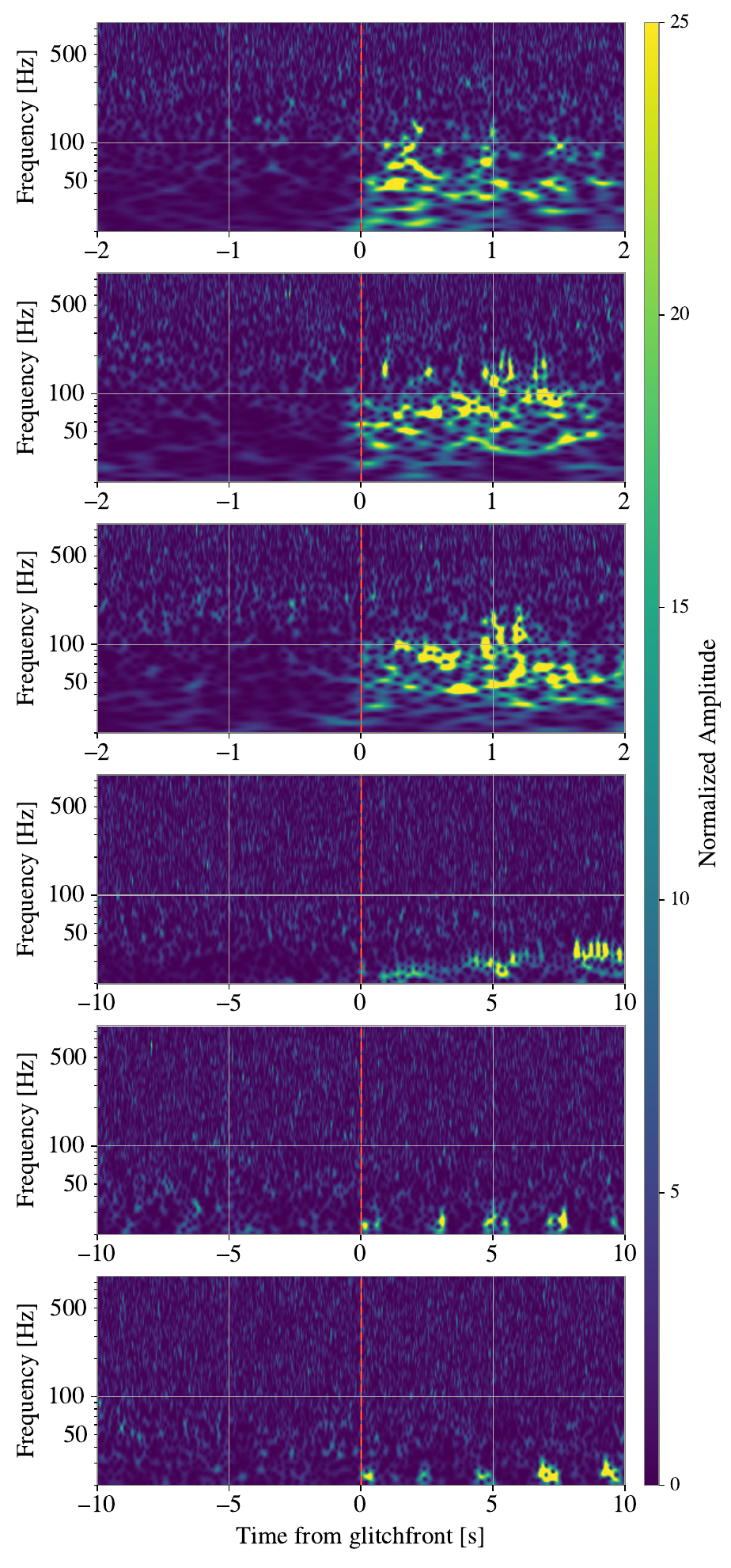}
    \caption{Omegascans of each thunder (top three plots) and fast-scattering (bottom three plots) glitch example, centered on the early glitch behavior. Glitch-fronts are found by selecting the earliest rectangular region in time-frequency space with SNR $\geq$10 in 1/3 of its bins, with rectangle dimensions dependent on glitch type. Red-dotted lines indicate the measured glitch-front.}
    \label{glitchfronts}
    \end{figure}

We then evaluate the minimum length of time between the ``chirp'' of our injected signal and start of each glitch at which we see significant variations in cost function measurements. The procedure to find the GPS time corresponding to the beginning of each glitch example is explained in the next section. Using this we obtain estimates for how close in time a signal and glitch can occur without significantly biasing the posterior distributions for certain parameters.

\subsection{Determining glitch-front times}
\label{finding_glitchfronts}

In Section \ref{safe_times}, we evaluate how close in time our CBC signals and glitch examples can be before the glitches begin to bias PE for the signals. In order to determine when the glitch begins, we need the ability to distinguish consistent, reproducible ``glitch-fronts'' between examples of the same glitch type. Glitch-fronts are used to calculate the earliest injection time relative to the glitch before we see significant bias in the posterior distributions.

Glitch types observed in GW detectors have a wide variety of characteristic time-frequency morphology, making it challenging to detect glitch-fronts with a single method. The glitch types chosen for this study have characteristic durations ranging from a fraction of a second to several minutes, while the characteristic frequency ranges span from 20 Hz to greater than 500 Hz. As such, we measure the glitch-front for each glitch type using a different method.

Glitch-fronts for thunder and fast-scattering examples used in this study are calculated from spectrograms of the GW strain data, referred to as ``omegascans'' (see Figure \ref{glitchfronts}). Glitch types are assigned rectangles in time-frequency space with width and height corresponding to characteristic duration and frequency range of the glitch-features, respectively. Iterating sequentially in time and then frequency, we move the rectangle over the omegascan and count the time-frequency bins within its area. When each time column within the rectangle has at least 1/3 of its frequency bins at SNR~$\geq$~10, the left-most point in the rectangle is classified as the glitch-front.

For thunder glitches, the characteristic time-frequency dimensions we use are 0.1 seconds and 20~Hz. For fast-scattering glitches, time intervals between the sub-second transient arches can vary, so the time-frequency rectangle dimensions are chosen to pick out individual arches from the signal. This corresponds to dimensions of 25 milliseconds and 1.5~Hz. Performing this calculation produces the red dotted lines in Figure \ref{glitchfronts}.

\begin{figure}
    \centering
    \includegraphics[width=0.8\linewidth]{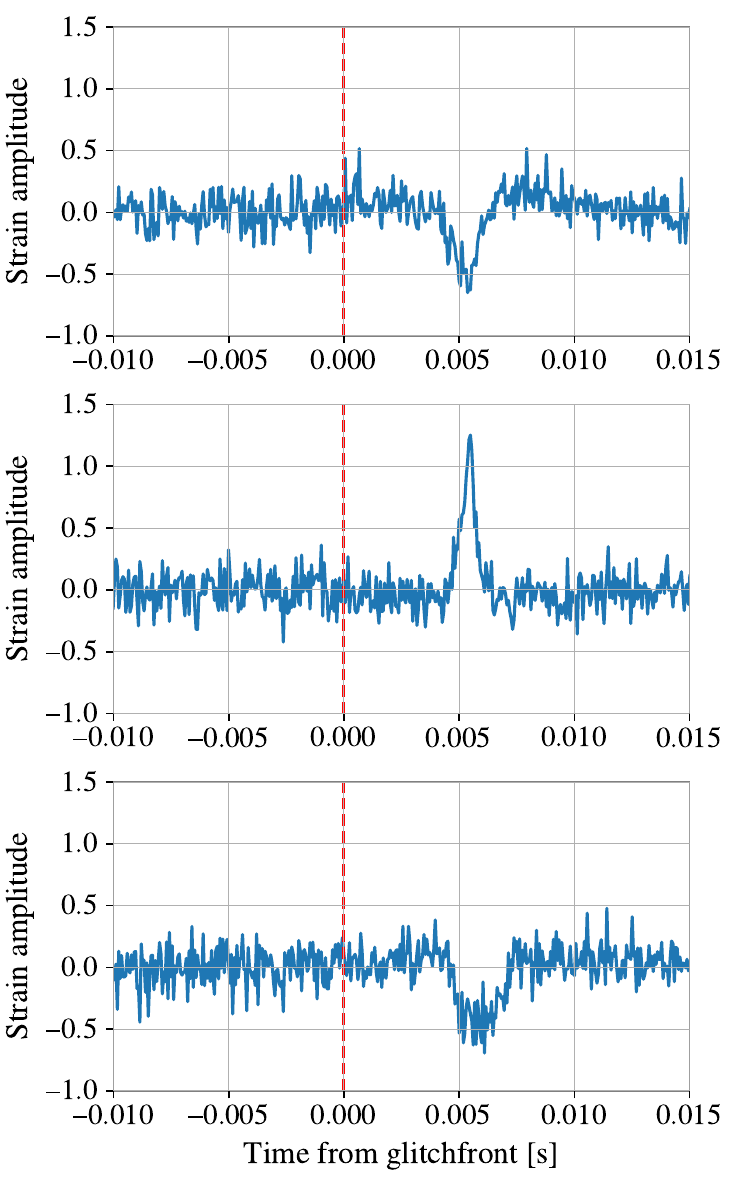}
    \caption{Timeseries of the LLO blip glitch examples used in this study. Glitch-front are found by measuring the earliest point 4$\sigma$ above the mean and subtracting 5 milliseconds. Red-dotted lines indicate the measured glitch-front.}
    \label{blip_fronts}
    \end{figure}

Blip glitches occur with higher SNR, which can cause them to appear to have longer durations in omegascans due to the filtering involved. As a result, the glitch-front finding method used for thunder and fast-scattering glitches detects high SNR regions in the data earlier than they actually occur. To estimate glitch-front times for blips, we measure the mean and standard deviation of the strain data timeseries prior to the glitch. We find the point at the which the strain signal exceeded four standard deviations (which should be well within the glitch) and then we set our glitch-front time as being 5 milliseconds before that point, slightly above the characteristic duration of a blip glitch. The glitch-fronts measured using this method are shown in Figure \ref{blip_fronts}.

\section {Results}

 We present our results in four sections: the maximum median deviations observed for biased mass, spin, and luminosity distance posteriors (Section \ref{median_devs}), trends in sky position posterior distributions (Section \ref{hpd_intervals}), the minimum ``safe'' separations in time between glitch and CBC signal (Section \ref{safe_times}), and glitch-tracking behavior (Section \ref{glitch-tracking}).

In this and subsequent sections we will discuss estimated parameters that have a statistically significant fraction of their posterior distribution probability density occurring far from the injected value, which we refer to as being ``significantly biased.'' To determine whether a posterior distribution is significantly biased, we compare its cost function value and standard deviation to reference cost and standard deviation thresholds which we determine from the glitch-free data. 

The expected distributions of our chosen metrics in Gaussian noise are not well-understood, and can vary between parameters. Assuming Gaussian-distributed posteriors for injections into glitch-free data, the standard error on our sample of standard deviations should be drawn from a chi-squared distribution. However, the scaling and position of this distribution will change for different parameter injection values, with each glitch-free injection set consisting of only 16 measurements. Furthermore, some parameters are bounded (for example: mass ratio is bounded by the primary and secondary mass values), making the assumption of Gaussian-distributed posteriors not necessarily appropriate. This uncertainty carries forward into the cost function, which is dependent on both the median distance from the injected value and the standard deviation. 

To mitigate this uncertainty, we set our thresholds using the extrema of the cost and standard deviation values in the control group posterior distributions. For each glitch, signal, and parameter, we calculate each metric's mean and standard deviation from the inner 75th percentile of the data, then calculate how many standard deviations away the minimum and maximum control group outliers are. We then take the global extrema measured for each parameter and used them as our metric thresholds. Posteriors outside of the control group that have costs and standard deviations outside of these thresholds are considered significantly biased. This thresholding method limits the possibility that biases deemed significant would be seen in ``clean'' control group data. From our calculated thresholds, we note that for all parameters the maximum control group cost outliers are much further from the 75th percentile mean than the minimum cost outliers, indicating heavier tails towards higher values. Standard deviation thresholds trend low or high depending on the parameter, which may be caused in part by prior range limitations.

We evaluate the quality of our control group posterior distributions to ensure that they could be used to determine the significant bias cost threshold. Within the control group posterior distributions there are two injection instances showing biases that far exceeded other injections. The credible intervals for the biased posteriors associated with these two injections show $<5\%$ support for the injected value. Given the susceptibility of short-duration signals to interference and the low deviation rate amongst control group injections (2 out of 432), the sources of the deviations are labeled as fluctuations in the Gaussian noise background, or perhaps short low-SNR glitches. These two potentially problematic injections are excluded from our results.

Additionally, it is worth noting that the threshold for deciding when bias is significant may exceed the maximum possible cost function and standard deviation values if the control group metrics are particularly noisy. In this case we assume that the associated posterior distributions are too poorly constrained to characterize significant bias. 

\subsection{Significant posterior distribution deviations}
\label{median_devs}

\begin{figure*}
    \centering
    \includegraphics[width=0.8\linewidth]{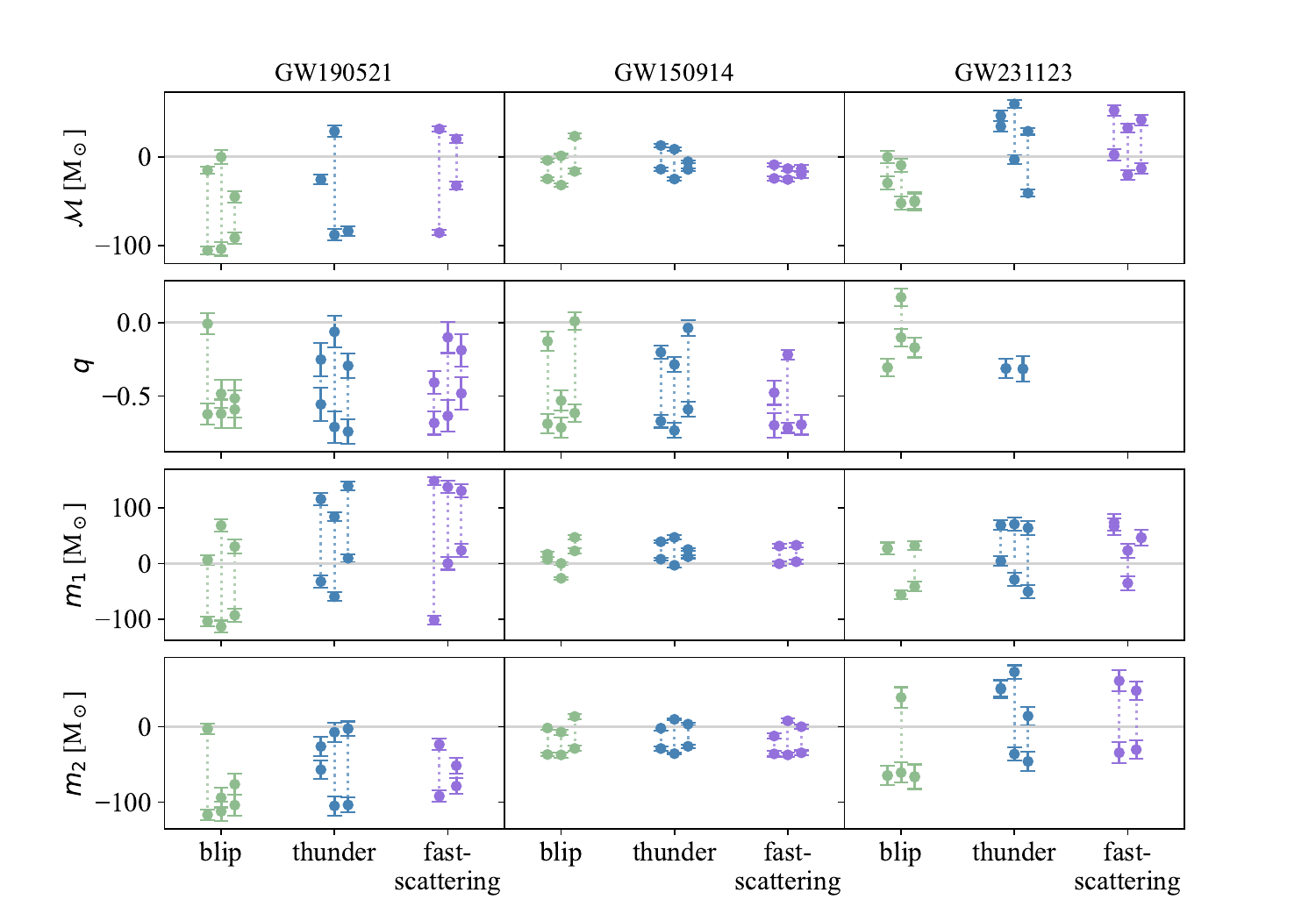}
    \caption{Maximum median deviations in observed mass parameters for each combination of signal and glitch, relative to the injected value. For a particular choice of glitch/signal/parameter, this shows the maximum range of possible biases to the median that we see in a three-detector network. Posterior distributions used in this plot exceeded cost and/or standard deviation thresholds for significant bias. Each connected pair of points above a glitch type corresponds to one of the three examples of that glitch type. Circular points on either side of a vertical line correspond to the minimum and maximum median deviations observed. Error-bars show one standard deviation on the control group medians.}
    \label{mass_medians}
    \end{figure*}

\begin{figure*}
    \centering
    \includegraphics[width=0.8\linewidth]{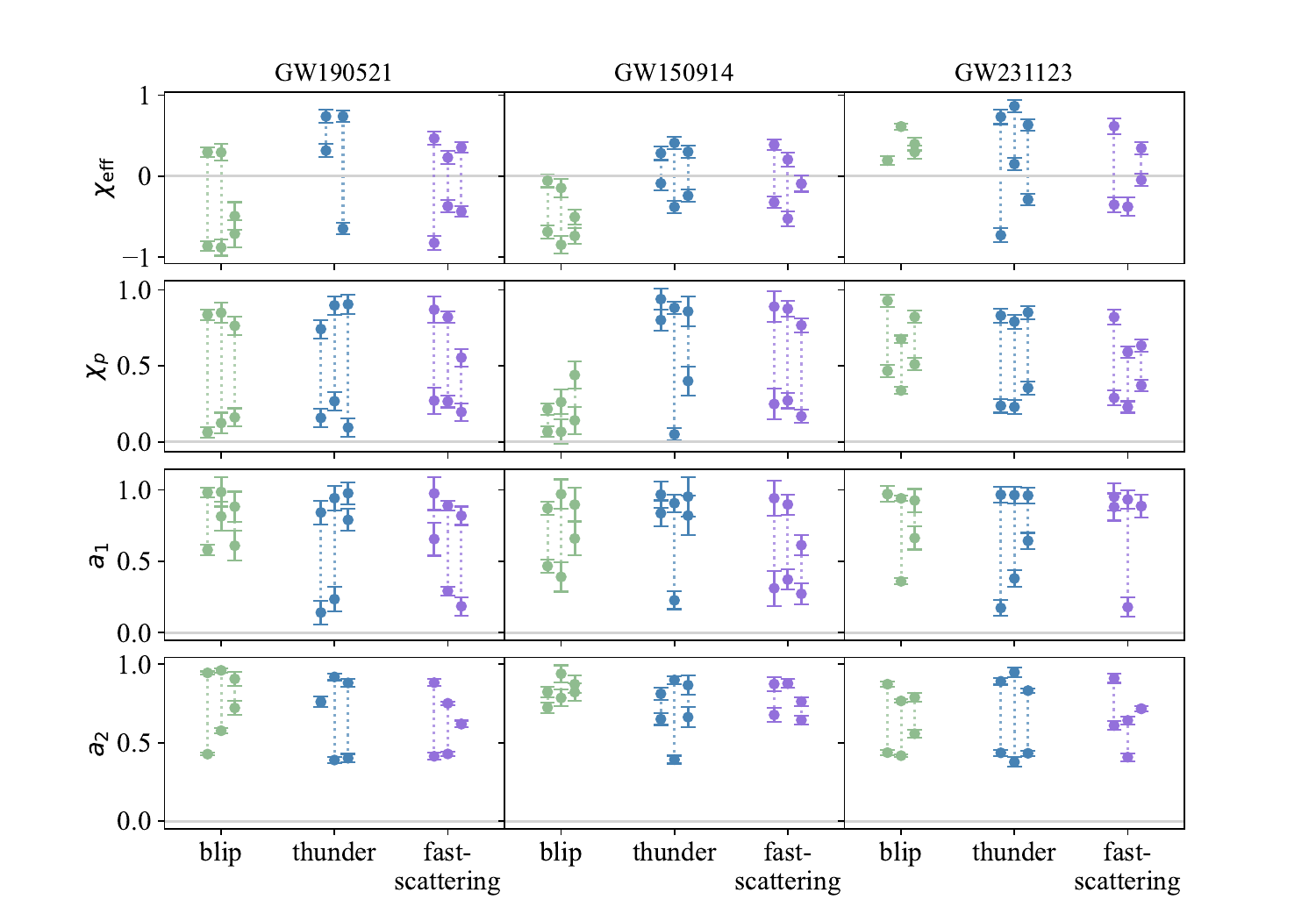}
    \caption{Maximum median deviations matching setup in Figure \ref{mass_medians}, but showing results for spin-related parameters. All injected signal templates have zero spin, meaning that $\chi_p$, $a_1$, and $a_2$ can only be biased towards higher spin values.}
    \label{spin_medians}
    \end{figure*}

\begin{figure*}
    \centering
    \includegraphics[width=0.8\linewidth]{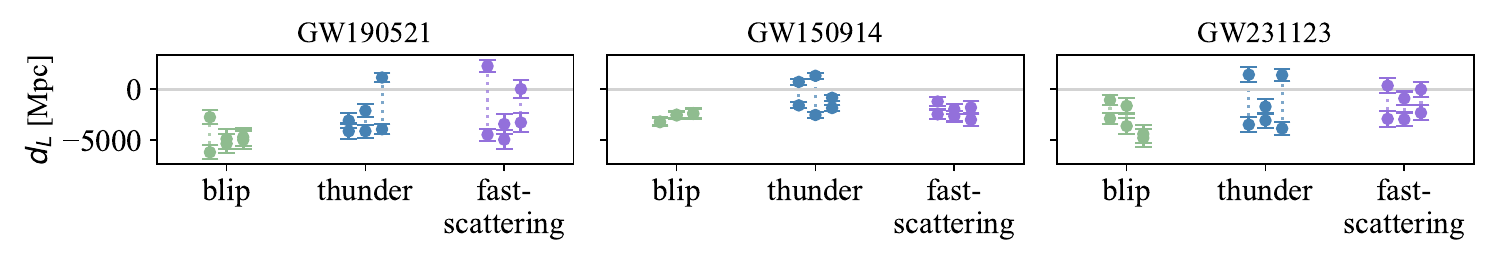}
    \caption{Maximum median deviations matching setup in Figure \ref{mass_medians}, but showing results for the luminosity distance ($d_L$).}
    \label{extrinsic_medians}
    \end{figure*}

 We present the range of median values observed in this ``worst-case scenario'' for interacting CBC signals and glitches. Median deviations reported in this section are restricted to injection times outside of the 16 control group measurements. The range of median deviations for parameters related to mass is shown in Figure \ref{mass_medians}. The y-axis shows the difference between the median value and the injected value, which is denoted by the light gray line at zero. Connected pairs of points indicate the minimum and maximum observed posterior median offsets for a single glitch example, parameter, and injected CBC signal. All points shown in Figures \ref{mass_medians}-\ref{extrinsic_medians} exceed bias thresholds, indicating that extrema values near zero likely have significantly different standard deviations than the control group. The error bars give the 1$\sigma$ standard deviation on the cost for the control group corresponding to that glitch example. Missing points indicate that one or zero significant biases were measured for that injection set. This plot shows the largest expected biases for the mass-related posterior distributions, from which we can constrain future glitch interference effects in short-duration CBC signals.

From this figure we see that almost all mass parameters show extreme bias as a result of interference from our selected glitch types. Excluding mass ratio measurements for GW231123-like injections near fast-scattering glitches, at least one example of each glitch type shows significant biases. In our more extreme cases we can see deviations of up to 100 M$_\odot$ in chirp mass and secondary mass. We also see in our study that, for some mass parameters, the median biases are uniform. This includes mass ratio for GW190521-like and GW150914-like signals, which are uniformly underestimated, and primary mass for GW150914-like signals interacting with thunder glitches, which are uniformly overestimated. None of the primary mass components are biased to the point of some part of the posterior falling within the neutron star range ($<3 M_\odot$).

The absence of mass ratio biases for GW231123-like injections near fast-scattering glitches implies invariance to interference, but is likely due to control group data quality. We see large variance in control group cost and standard deviation values for mass ratio estimates on GW231123-like signals across all glitch types, indicating that the mass ratio for this signal is susceptible to fluctuations in the background noise. That being said, these preliminary results indicate that these biases are smaller for fast-scattering glitches than for blip and thunder glitches.

There is one example where the maximum bias range is small, and consistent across all three glitch examples. The chirp mass posterior biases for the GW150914-like signal interacting with our fast-scattering glitches shows maximum deviations with a range of 5 to 30 M$_\odot$. These biases, while large relative to the true chirp mass of the CBC signal (28 M$_\odot$ in source-frame), are also small relative to those caused by the other glitch types, indicating predictable behavior for that subset of our parameter estimation data.
 
Parameter biases are also present for all glitch types, signals, and spin-related parameters, as shown in Figure \ref{spin_medians}. Similar to the previous plot, we see that all spin parameters show significant bias due to glitch interference. In particular, we see several instances of near-maximal spin parameters for systems injected with zero spin, showing that low-duration CBC spin parameters have extremely high susceptibility to interference from the selected glitches types (as previously reported by previous studies \cite{Ashton_2022}, \cite{shah_2023}, \cite{powell}, \cite{Abbott_2018}). Given the fact that the injected values for precessing spin and both component spins are zero, there is no possibility of underestimation of their posterior distributions. Effective spin posteriors show bias towards both aligned and anti-aligned spins in most cases, with only the biases between blip glitches and GW150914 and GW231123-like signals being consistently in one direction.

Similarly, we present median deviations for the luminosity distance of the event (Figure \ref{extrinsic_medians}). Luminosity distance shows significant bias for all combinations of glitch/CBC signal, with the most extreme cases being biased by over 5000 Mpc. Several combinations of glitch and signal are exclusively negatively biased, which likely results from the glitch SNRs generally being significantly higher than the CBC signal SNRs. Other extrinsic parameters, such as sky position and geocenter time, are excluded due to their inherent multimodality, which produces uninformative control and standard deviation values. In Section \ref{hpd_intervals} we investigate the credible interval support for the true sky positions, and in Section \ref{glitch-tracking} we explore biases in geocenter time.

Overall, we find bias in all parameters for each CBC signal template and glitch type used in this study. These signal templates cover a wide range of component masses, indicating that short duration signals in general are susceptible to interference from our glitch types. With regards to event validation, this means that \textbf{\textit{short duration CBC signals overlapping the glitch types used in this study should be treated as having significant biases in mass and spin parameters and luminosity distance. These cases of glitch interference will require glitch subtraction}}. The exact time-separation between glitch and CBC signal at which biases occur is explored more in Section \ref{safe_times}.

Many of the significant posterior biases we measure are biased \textit{towards} the injected value. This finding is in agreement with Udall et al, which found instances of glitch overlap resulting in more accurate spin posterior distributions \cite{Udall_2025}. These will often have cost values that do not exceed our threshold, but standard deviations that fall below our standard deviation threshold. For posterior distributions flagged as being significantly-biased in cost or standard deviation, 63.8\% have standard deviations below the threshold. By contrast, 8.7\% of biased posteriors result in standard deviations above the maximum standard deviation threshold, and 27.4\% are within the threshold boundaries. This indicates that, for our set of relatively low-SNR CBC signals and high-SNR glitches, \textbf{\textit{posteriors are more likely to be biased towards narrower distributions than wider distributions}}. As a result, PE conducted on signals contaminated by nearby glitches could likely yield spurious constraints on black hole spins.

All biased mass, spin, and extrinsic parameter posterior distributions exhibit majority decrease in standard deviation. Spin components show a much higher preference towards decreasing standard deviations, likely due to the component spins and precessing spin being largely uninformative and therefore spanning the full prior range in the control case. As previously mentioned, biased luminosity distance posteriors tend heavily towards low standard deviations.

It is unclear in most cases to what extent the parameter estimation is fitting the glitch instead of the CBC signal. A previous study by Ashton et al (2022)~\cite{Ashton_2022} provides context for the trends we see in the maximum median deviations. They performed parameter estimation on glitch examples to infer the parameter ranges spanned by CBC model fits of different glitch types. This study shares two glitch types with our own, blip and fast-scattering. The inferred properties for these glitch populations show a tendency towards more extreme mass ratios and high primary component spins than would be expected from CBC signal populations. In Figure \ref{mass_medians} we see that the majority of our observed mass ratio biases tend towards smaller values. For the primary and secondary spin magnitude, we cannot bias our spins any lower than their injected values at zero, but in Figure \ref{spin_medians} we do see that the largest component spin biases in the blip and fast-scattering cases are near-maximal, which mirrors the glitch posterior distributions from the study.

\subsection{Credible intervals for biased sky position posteriors}
\label{hpd_intervals}

    \begin{figure}
    \centering
    \includegraphics[width=1.0\linewidth]{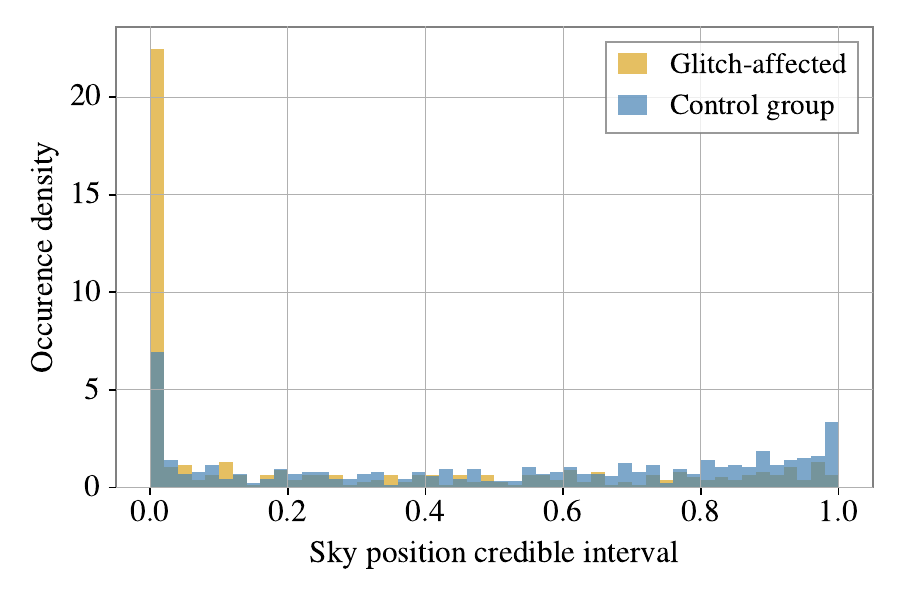}
    \caption{Histogram of sky position credible intervals for control group and glitch-affected posterior distributions. Glitch-affected sky position posteriors (yellow) are defined as having at least one other parameter corresponding to the same injection classified as significantly biased. Control group posteriors (blue) are taken from injections into Gaussian background data prior to the glitch.}
    \label{cred_intervals}
    \end{figure}

In this section we investigate changes to sky position posteriors by looking at support for the true sky position from the right ascension and declination posterior distributions. In contrast to the median offsets we looked at in the previous section, which indicate the degree to which the median has been biased, Highest Posterior Density (HPD) intervals give our confidence in the true value from a given posterior distribution. This is due to the fact that negligible support for the true sky position makes it unlikely for the true sky position to be located. We find that, of our 27 glitch/signal combinations, only one shows $>$0.3\% support for the true sky position for the entirety of its non-control group injections. These correspond to one blip example and the GW231123-like signal. For the control group, sky position posteriors showed $>$5\% support for the true sky position for all injections for 14 out of the 27 injection sets and $>$0.3\% support for all injections in 21 injection sets. This indicates that the \textbf{\textit{sky position posteriors are also susceptible to significant bias due to interference from all glitch types used in this study}} Sky position posterior distributions for BBH signals overlapping with glitches are likely to have little to no support for the true sky position, which will affect follow-up analyses by light-based observatories. 

We also report the rate of sky position posterior biases between control group and glitch-affected injections. A histogram of sky position credible intervals for both cases is shown in Figure \ref{cred_intervals}. Glitch-affected sky position posteriors are defined as having at least one other parameter corresponding to the same injection classified as being significantly biased. We can see that glitch-affected injections are approximately four times as likely to have sky position posteriors with little to no support for the true sky position. The credible interval measurements with less than 20\% support for the true sky position make up 27.5\% of the total measurements in the control group posterior set and 58.7\% of the total measurements in the glitch-affected posterior set.

These results can be compared to those from a similar study by Macas et al. (2022)~\cite{macas}, which investigates low-latency GW event localization in the presence of detector glitches. This study used the same glitch types as our study, and shares two of our CBC signal templates, GW190521 and GW150914. Macas et al. used field-of-view (FOV) tiles, which are the numbers of regions that a telescope would need to observe before seeing the true sky position. Similarly to the credible interval method we use, the tiles are counted from regions of higher probability density to smaller probability density. Using the low-latency detection package \texttt{PyCBC Live}~\cite{Dal_Canton_2021}, they found that thunder and fast-scattering glitches in three-detector networks do not have significantly-biased sky localization. They also found that blip glitches do not cause sky position biases unless they are within 0.03 seconds of the signal chirp, while our own blip glitch examples show sky position biases within 0.1 seconds of the signal chirp (see next section). Comparing our results with Macas et al. implies that the low-latency sky localization method used by \texttt{PyCBC Live} may be more robust to glitch-bias than Bayesian parameter estimation. That being said, it is worth noting that the injection SNRs used in Macas et al. were higher than those used in this study, with network SNRs ranging from 27-41.

\subsection{Deviation times relative to glitch front}
\label{safe_times}

We investigate the earliest times relative to the glitch-front that significant posterior biases occur to determine when potential glitch interference can be safely ignored. For each combination of glitch and signal, we select the earliest injection in which significant bias was observed and compare it to the glitch-front time. These injection periods are limited to be within 2 seconds prior to the glitch-front. This is due to the fact that our parameter estimation models only use data spanning the CBC signal, with a time-prior range of 0.1 seconds on each side, plus 2 seconds after the signal chirp for ringdown. If the injection occurs more the 2 seconds prior to the glitch-front, no data from the glitch will be included in the parameter estimation.

\begin{figure}
    \centering
    \includegraphics[width=1.0\linewidth]{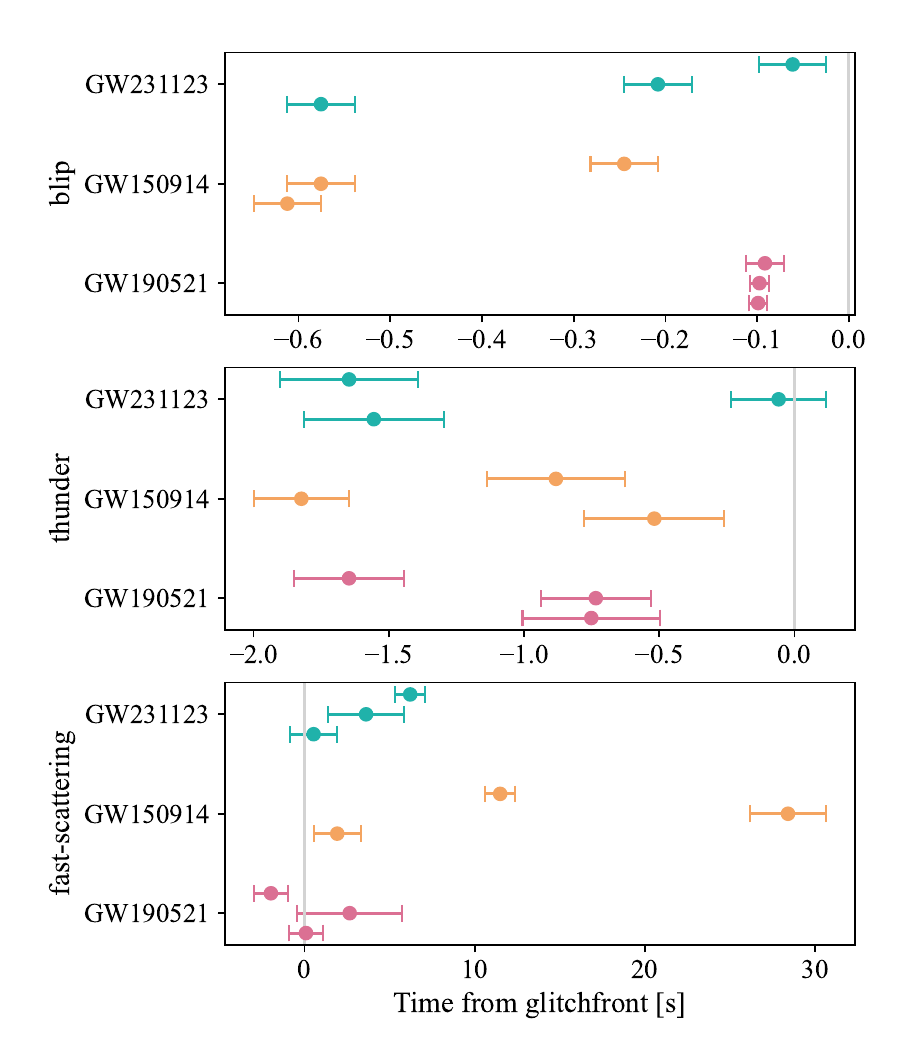}
    \caption{Time of first biased posterior relative to the beginning of each glitch. Times are reported such that negative values indicate bias prior to the glitch-front and positive values indicate bias after. The glitch-front is indicated by the gray line. Error-bars indicate the time intervals between injections for that glitch example. Biased posteriors are limited to be within 2 seconds prior to the glitch-front, as glitch data is not present in the analysis otherwise. Note the varying timescales for each plot.}
    \label{safe_times_plot}
    \end{figure}

The results of this are shown in Figure \ref{safe_times_plot}. Times are reported relative to the glitch-front, such that negative values indicate the bias occurring prior to the glitch-front and positive indicate the bias occurring after. We see that for all signals interacting with blip and thunder glitches, significant posterior biases occur prior to the glitch-front. Several of these biases occur outside of the time prior. For fast-scattering glitches, the majority of posterior biases are observed after the glitch-front, indicating that the early behavior of our fast-scattering glitch examples do not significantly impact our signals.

A study by Hourihane et al. (2025)~\cite{Hourihane_2025} finds that posteriors are not significantly impacted by glitches with SNR below 50 unless they occur within the time prior and characteristic frequency of the CBC signal. This study simulates each component of the data: the CBC signal, the glitch, and the Gaussian background noise. Similar to our study, they also focus on relatively high detector-frame mass events ($m$ $\in$ (20, 100)~$M_\odot$), but the range of masses used falls below our maximum detector-frame mass of 249~$M_\odot$ for our GW231123-like signal. The glitches used are individual simulated sine-Gaussian Morlet-Gabor wavelets which, when summed, are expected to be able to approximate most glitch types seen in LIGO-Virgo detectors\cite{bayeswave}. 

This study offers a unique point of comparison to our own. While our detector-frame component mass variability is smaller, our range of masses used is larger. Additionally, in Hourihane et al., the Morlet-Gabor wavelets and simulated Gaussian noise used are assumed to reproduce real noise seen in GW detectors, but there are key uncertainties in those assumptions. The full range of glitch morphology has not been modeled using Morlet-Gabor wavelets and, if there are glitches that are not well-approximated by these wavelets, they would have more residual post-subtraction noise and could potentially cause larger biases than anticipated. Additionally, LIGO-Virgo background noise exhibits non-Gaussian behavior which may interfere with the background assumed by the parameter estimation model\cite{Zackay_2021}. 

At first glance, the results of Figure \ref{safe_times_plot} appears to disagree with the Hourihane et al. study. However, deeper investigation into our results reinforces the assertion that glitch biases are most significant when they occur within the CBC signal time prior.

We first look at the prevalence of biases that occur as a function of the proximity between the CBC signal and glitch. We separate our significant bias instances into three categories based on whether there was glitch power within the signal's time prior, outside of the time prior but within the signal's 2-second ringdown window, and outside of the ringdown window. From this we find that 98.1\% of biases exceeding the cost thresholds and 97.3\% of biases exceeding the standard deviation thresholds occur within the time prior, with the remaining bias instances being split relatively evenly between the in-ringdown and outside-ringdown categories. The actual ratio of data within the time prior versus outside of it is 52.9\%.

We can also compare the severity of biases seen in each category. We see that the average above-threshold cost value for biased posteriors within the time prior is 20.5 times larger than the same bias for instances in the in-ringdown category, and 23.2 times larger than those in the outside-ringdown category. Similarly, the average below-threshold standard deviation for biased posteriors within the time prior is 4.3 times larger than the in-ringdown category and 3.1 times larger than in the outside-ringdown category. The average below-threshold cost and above-threshold standard deviation biases are not significantly different between the three categories. Combining this with the previous paragraph we see that \textbf{\textit{posterior biases are both more common and more severe when there is glitch energy within the time-prior compared to outside of it}}.

The difference in results is likely due to the metrics used in both studies. By using fully-simulated data, Hourihane et al. are able to directly compare posteriors for identical data with and without the glitch. The metrics used (Jensen-Shannon divergence and distribution reweighting) have known properties that allow for statistically-motivated thresholding for significant bias. In our study we use the range of control group cost and standard deviation values to create threshold boundaries. This method assumes that the range of glitch-free cost and standard deviation values for our parameters is described completely by our control group data, but this is limited by the number of glitch-free posteriors measured per parameter. We find that by increasing the cost thresholds by a factor of 1.9 and standard deviation thresholds by a factor of 2.4, the remaining biased results all fall within the time prior of the CBC signal.

The safe time intervals we observe can also be compared to the results of Ghonge et al. (2024)~\cite{Ghonge_2024}, which looked at CBC signal posterior biases caused by glitch interference before and after glitch-subtraction. This study injected a simulated signal that mimics GW150914 at different points in time relative to glitch examples from three different glitch types, one of which (blip glitch) is shared with our study. They performed parameter estimation before and after glitch-subtraction, looking specifically at chirp mass, mass ratio, and effective spin. They found that blip glitches cause moderate biases to their estimated parameters, and that these biases occur when the peak SNR region of the blip is within 0.1 seconds of the CBC signal. The degree of bias observed by Ghonge et al. is similar to what we observe in Section \ref{median_devs}, but is slightly smaller on average. This may be affected by the fact that the CBC signal SNR was fixed at 15, 3-5 higher than the SNRs used in our study. Additionally, the direction of bias is generally the same, favoring lower chirp mass, mass ratio, and effective spin. Lastly, we see that the safe time interval used by the study lines up with the conclusion of our study and Hourihane et al.: that severe biases occur primarily when there is glitch energy within the time prior of a CBC signal.

\subsection{Glitch-tracking behavior}
\label{glitch-tracking}

\begin{figure}
    \centering
    \includegraphics[width=1.0\linewidth]{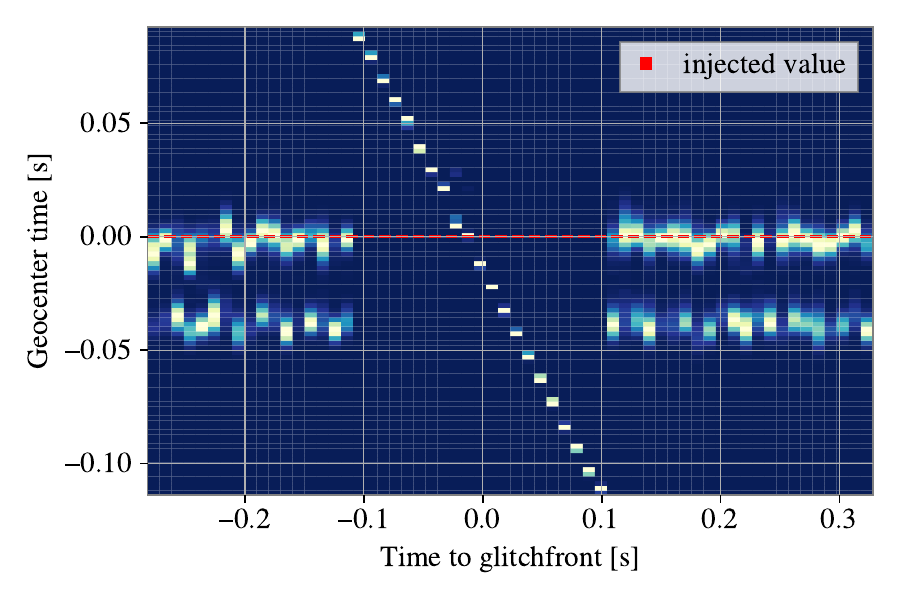}
    \caption{Geocenter time posterior distributions for a GW190521-like signal injected at different points in time relative to a blip glitch. The  vertical axis indicates the geocenter time posterior minus true geocenter time. The red-dotted line shows the true value. Within the 0.2 second window where the glitch is within the time prior, the posterior distributions follow a near-diagonal line tracking the glitch.}
    \label{glitch_tracking}
    \end{figure}

We lastly describe ``glitch-tracking'' behavior for blip glitches interacting with GW150914-like and GW190521-like signals. We define glitch-tracking as instances when the parameter estimation model attempts to remove the glitch from the data instead of the signal. For all blip glitch examples overlapping both our GW190521-like and GW150914-like signals, the geocenter time posterior distributions center around the blip for the entirety of the time that the glitch is within the time prior of the CBC signal. An example of this can be seen in Figure \ref{glitch_tracking}. Around the -0.1 second mark, the glitch enters the time prior of the signal and the posterior distribution leaps forward to where the glitch is in time. It then tracks it backwards with every time interval, finally resuming its normal behavior when the glitch falls outside the time prior near the 0.1 second mark. This indicates that the maximum median deviations seen in Figures \ref{mass_medians}-\ref{extrinsic_medians} for these glitch/signal combinations may be the result of the parameter estimation model fitting to the glitch itself, rather than either the signal or a combination of the signal and glitch.

The parameter estimation model assumes a gravitational-wave signal and stationary Gaussian noise only, and adjusts its model parameters to minimize the difference between the residuals and a Gaussian background. When the glitch and signal directly overlap, we expect the model to fit the combined data to minimize excess noise; however, the circumstances under which the model prefers the glitch over the signal are less-well understood. The glitch-tracking we observe indicates that our waveform model is able to fit our blip glitch examples well enough that the remaining GW190521-like and GW150914-like signals are more consistent with the Gaussian background noise.

Interestingly, we do not see this behavior as strongly for blips glitches overlapping our GW231123-like signal. Glitch tracking is observed for only one blip example, and not for the entirety of its presence in the time prior of the signal. This implies that \textbf{\textit{there may be some feature of our GW231123-like signal that makes it more robust to nearby blip glitches}}. One feature of GW231123 that may affect its robustness is that it has a much lower frequency overlap with our blip glitch examples than our other signal templates. Additionally, it is possible that glitch-tracking can occur with the other glitch types used in this study, but due to the time intervals used to iterate over the thunder and fast-scattering glitches being on the order of or larger than the time prior, we are unable to assert this with the same level of certainty.

The median biases we observe during periods of glitch-tracking likely correspond to the modeling of the glitch on its own. This is reinforced by the Ashton et al. (2022) study that found that estimated posterior distributions for blip glitches tend to have low mass ratios and high primary component spins~\cite{Ashton_2022}.

\section{Conclusion}

In this study we investigate the range of possible parameter estimation biases that detector glitches can cause when interacting with short-duration CBC signals. The detector in which the glitches occur and SNR of the simulated CBC signal are chosen to obtain worst-case scenario estimates of parameter estimation biases for a three-detector network with otherwise good data quality. From our estimated parameters we explore the maximum posterior distribution biases, cases where parameters may be robust to glitch interference, and what constitutes a safe time separation between the CBC signal and the start of a glitch, where no glitch subtraction is required prior to parameter estimation

Our study has shown that \textbf{\textit{short duration CBC signals are extremely sensitive to parameter estimation biases from the three glitch types used}}. Biases occur in all parameters for a majority of signal and glitch types. Some parameters are biased consistently in one direction, indicating that we may be able to predict the type of bias for certain glitch-affected signals. However, the range of possible deviations tends to be large, and can change between different examples of the same glitch, making the degree of bias challenging to determine without glitch subtraction. 

We similarly see that \textbf{\textit{all glitch types have the potential to significantly bias sky position posteriors for the CBC signals used in the study}}. CBC signals that had at least one other biased parameter were more likely to have low posterior support for the true sky position. We also note that another study by Macas et al. (2022)~\cite{macas} that focuses on low-latency sky localization for CBC signals in the presence of glitches reports evidence that low-latency sky maps are significantly more robust to glitch interference than our full parameter estimation methods.

Safe time separations between the signal and glitch are measured and compared to the results from Hourihane et al~\cite{Hourihane_2025}. We find that our control group-based thresholding method resulted in significant biases being detected when all glitch power is outside of the CBC signal time prior, but that a small increase in our threshold values restricts all above-threshold biases to be within the time prior. Additionally, we show that biases that occur within the time prior make up a significant fraction of the total biases seen, and that the degree of bias is on average much larger for glitches within the time prior than outside of it. Thus we find that our results agree with Hourihane et al. (2025), providing more evidence that \textbf{\textit{parameter biases are more severe when glitch energy is present within the time prior of the CBC signal}}. Our study verifies this for a wider range of component masses and for real strain data and glitches.

Lastly, we investigate instances where the parameter estimation model appears to be modeling the glitch rather than the signal. This behavior is limited to blip glitches with GW190521-like and GW150914-like signals, with GW231123-like signals showing a small amount of glitch-tracking. The signal features that make GW231123 more robust to glitch-tracking behavior require further study.

There are several ways this study can be built upon in the future. While short duration CBC signals are expected to be the most susceptible to glitch interference, the full effects of glitches on parameter estimation for longer duration signals is not well-understood. Similar studies could be performed on lower mass binary black hole and neutron star signals. Additionally, a larger variety of glitch types could be used to make the results more broadly applicable.

Future studies could also benefit from simulating each aspect of the data rather than just the CBC signal, as was done in Hourihane et al~\cite{Hourihane_2025}. One major limiting factor in this study is finding glitch examples that are representative of the larger population, have at least 64 seconds of glitch-free time prior to the glitch-front, and are glitch-free in the other two detectors during this time window. Using simulated background noise and glitches would allow us to more fully test possible interactions between the glitch and the signal. It would also provide the ability to look at the posterior distribution for a CBC signal with and without a glitch present. Differences between the distributions could be directly attributable to the glitch, without any uncertainty regarding Gaussian background effects on parameter estimation. While glitch modeling is only available for a subset of glitch types, this framework would make subsequent studies more efficient and informative.

In conclusion, we find that mass, spin, and extrinsic parameters investigated were all susceptible to interference from nearby glitches. These biases are most severe when power from the glitch was within the time prior of the CBC signal. We also see instances where the parameter estimation model appears to prefer the glitch to the CBC signal. While this study focuses on a few glitch classes, CBC signals, and injected parameter values, it provides preliminary estimates for how these factors interact to guide future analyses.

\section{Acknowledgements}

Special thanks to James Clark and Sylvia Biscoveanu for their assistance with submitting computing jobs through the IGWN Computing Grid. Thanks to Laura Nuttall, Josh Pooley, Vivien Raymond, and Rory Smith for contributions to earlier versions of this project. Additional thanks to Gregory Ashton for work as internal reviewer and Meg Millhouse for feedback given during interval review process. This project was funded in part by the NSERC Discovery Grant and the McDonald Institute. This material is based upon work supported by NSF's LIGO Laboratory which is a major facility fully funded by the National Science Foundation. The authors are grateful for computational resources provided by the LIGO Lab and supported by National Science Foundation Grants PHY-0757058 and PHY-0823459, as well as the computational resources provided by the Digital Research Alliance of Canada. SJM and KC were supported by NSF Grants PHY-2308770 and PHY-2409001.

\section*{Appendix A: Single template matched filter SNR recovery}
\label{matchedfilter}

This description follows Usman et al. (2016)~\cite{pycbc}. Single template matched filter recovery is a method of calculating the conformity of GW strain data to a single injection template. This is done by obtaining a $\chi^2$ statistic between the strain data $\rho_i$ and the template $\rho$. The signal template is split into $p$ frequency bins with equal energy and the energy in each bin is compared to the excess energy in the data:

\begin{equation}
    \label{matched_template}
    \chi^2 = p \sum_{i=1}{p}\left[\left(\frac{\rho_{cos}^2}{p} - \rho_{cos, i}^2\right)^2 + \left(\frac{\rho_{sin}^2}{p} - \rho_{sin, i}^2\right)^2 \right]
\end{equation}

\noindent where $(\rho_{cos}$, $\rho_{sin})$ and $(\rho_{cos, i}$, $\rho_{sin, i})$ are the orthogonal phases of the matched filter and the data, respectively. Higher $\chi^2$ values correspond to a higher likelihood that the signal is a noise transient. For $\chi^2$ values less than 1, the strain data is considered to be a good match to the template and the matched-filter SNR is given as the SNR of the data $\hat{\rho} = \rho$. For $\chi^2$ values greater than 1, the matched filter SNR is decreased by the function: 

\begin{equation}
    \hat{\rho} = \frac{\rho}{[(1 + (\chi^2_r)^3)/2]^{1/6}}
\end{equation}

\noindent where $\chi^2_r = \chi^2/(2p-2)$ is the reduced $\chi^2$. As a result, the matched filter SNR derived from the data gets reduced for having excess noise in its frequency bins. If this reduction in SNR causes it to fall below the detection threshold SNR, then the candidate is ignored in a PyCBC search.

In our study we inject signals with the same optimal SNR at various points of overlap with each glitch and use the true signal template to compute the matched filter SNR (see the second panel in Figure \ref{multi-panel}). Fluctuations/reductions in SNR are due solely to excess power from the Gaussian background and the glitch.  This output is used as a rudimentary metric for how much the glitch interferes with the signal. Additionally, by decreasing the injected SNR of the signal until the recovered SNR nears the minimum detection threshold, we can investigate ``worst-case scenario'' biases in parameter estimation, where the SNR is low but the signal is still confidently detectable.

\FloatBarrier

\bibliography{glitch-pe}

\end{document}